\documentstyle[12pt,aaspp4]{article}

\def\del{\partial}
\def\<{{\langle}}
\def\>{{\rangle}}

\def\sA{{\mathcal{A}}}

\def\sC{{\mathcal{C}}}
\def\sD{{\mathcal{D}}}
\def\sE{{\mathcal{E}}}

\def\ss{{{t^r}_\phi}}
\def\css{{t_{(r)(\phi)}}}
\lefthead{Gammie and Popham}
\righthead{Accretion in Kerr Metric}
\begin{document}
\title{Advection Dominated Accretion Flows in the Kerr Metric:
	I. Basic Equations}
\author{Charles F. Gammie\altaffilmark{1} and Robert Popham}
\affil{Harvard-Smithsonian Center for Astrophysics, MS-51 \\
60 Garden St., Cambridge, MA 02138}

\altaffiltext{1}{also Institute of Astronomy, Madingley Road,
	Cambridge CB3 0HA, United Kingdom}

\today

\begin{abstract}

We write down and solve equations describing steady state, optically 
thin, advection-dominated accretion onto a Kerr black hole.  The
mean flow, described by the relativistic fluid equations, is 
axisymmetric and vertically averaged.  The effect of
turbulence in the flow is represented by a viscous shear
stress.  Our treatment differs in several important ways from 
earlier work: we use a causal prescription for the shear stress, 
we do not assume the relativistic enthalpy is unity (this is 
important for rapidly rotating holes), and we use a relativistic
equation of state.  We present several representative solutions
and use them to evaluate the importance of relativistic effects,
to check our approximations, and to evaluate the robustness of
the input physics.  Detailed properties of the solutions are
described in an accompanying paper.

\end{abstract}

\section{Introduction}

The structure of accretion flows close to the horizon of a
black hole is of considerable astrophysical interest because:
(1) most of the gravitational binding energy is released close
to the black hole;
(2) there are strong-field gravitational effects close to 
the horizon that are unique to black holes;
(3) temperatures are highest close to the horizon, so exotic
physical processes may occur there.

Early studies of disk accretion onto black holes assumed a thin disk.
\cite{ss73} presented the basic equations used to describe thin disks,
but did not include relativity.  Accretion in thin, relativistic disks
was first described by \cite{bpt72} and \cite{nt73}.  Note that
recently \cite{rh96} corrected an algebraic error in NT's version of
the vertical structure equation.

\cite{pb81}, \cite{mp82}, and \cite{a88} introduced a set of disk
equations, dubbed the ``slim disk'' equations by \cite{a88}, which
include a number of terms neglected in the thin disk formulation.
These equations include radial pressure gradient, acceleration, and
energy transport, and provide the basis for the equations used in this
paper and most of the other recent work cited here on disk structure
around black holes.  These studies all used the ``Paczy\'nski
potential'', a pseudo-relativistic potential designed to reproduce the
main features of orbits around a nonrotating black hole (\cite{pw80}).

One of the terms included in the slim disk equations is the inward
advection of entropy by the accreting gas.  \cite{np93} showed that
this term, which is insignificantly small in the usual thin disk,
becomes large in the optically thin boundary layer region of
low-accretion rate cataclysmic variables.  Narayan and Yi recognized
the existence of a class of flows in which advection dominates the
energy balance, and found a self-similar solution in the
advection-dominated limit (\cite{ny94}, \cite{ny95a}).
Advection-dominated solutions were recognized as a new class of
thermal equilibria (\cite{a95}, \cite{c95}, \cite{ny95b}).
They resemble some earlier models by \cite{rbbp} and \cite{bm82},
where the hot accreting gas radiates very inefficiently.

In this paper we concentrate on a particular class of flows: optically
thin, advection-dominated accretion flows, or ADAFs. These flows cool
slowly, so most of the heat deposited in them by the dissipation of
turbulence (which also carries off angular momentum) is advected into
the black hole rather than radiated away.  The accretion is therefore
inefficient compared to a thin disk, which can radiate away a
substantial fraction of the accreted rest mass energy.  Over the past
few years, the theory of these flows has been developed and applied to
a number of observed black hole systems (see \cite{n97} for a review).
Most recently, \cite{nkh} (NKH) and \cite{cal97} presented global,
advection-dominated disk solutions in the Paczy\'nski potential.  Our
work differs from this in that it is fully relativistic, and includes
effects due to strong rotation of the black hole.

Recently, a number of authors have included the effects of relativity
in their models of disks around black holes.  \cite{l94} first wrote
down slim disk equations which included relativistic effects. This was
superseded by the work of \cite{acgl} (ACGL), who corrected some
errors in Lasota's work and provided sample numerical solutions.  Our
work differs from this one in many details, but mainly in our use of a
turbulent stress prescription that is explicitly related to the
Navier-Stokes stress and that is causal, and in our inclusion of the
relativistic enthalpy.  \cite{alp} (ALP) revisited the question of
vertical structure and derived a new expression for the scale height
beginning with the Euler equations.  We adopt ALP's version of the
vertical structure equations, with a minor correction.  Independently,
\cite{pa97} (PA) gave a very nice and rather complete treatment of the
problem, which included a causal viscosity and the relativistic
enthalpy; however, they used a polytropic equation of state rather
than solving the energy equation as we do.  \cite{ch96} has also
considered rotating accretion flows in the Kerr metric, but only in
the weak viscosity limit.  Also, unlike all of the above, we use a
relativistic equation of state.  This correction increases in
importance as the hole rotates more and more rapidly.  Finally, we
correct some minor errors in some of the above treatments.

The spirit of this work is to do the best possible job on the physics,
including the relativity and turbulent shear stress, within certain
limitations.  One limitation is physical: there is no well justified
formalism for treating the effects of turbulence in these flows.  We
treat the mean flow using the full equations of relativistic
hydrodynamics, and include the effects of turbulence via a turbulent
shear stress, making the minimal modifications of the inviscid
equations necessary to allow angular momentum transport and therefore
accretion.  Another limitation is complexity: we are forced to make
some approximations simply because without them the equations would be
too difficult to solve.  Despite these limitations, this work reveals
some generic properties of relativistic ADAFs, and, when coupled with
a proper scheme for relativistic photon transport, will allow us to
model their observational appearance.

The plan of this paper is as follows.  In \S 2 we explain our units,
notation, and record the relevant metrics and frames.  In \S 3 we write
down the basic dynamical equations.  In \S 4 we consider the heart of
the physics in this problem, the turbulent shear stress.  In \S 5 we
evaluate the critical point conditions.  In \S 6 we give a sample
solution (a full survey of solutions is described in \cite{PG}), and
evaluate our approximations.  \S 7 contains our conclusions.

\section{Preliminaries}

\subsection{Units and Metric}

We adopt $G = M = c = 1$ as our basic scalings, where $M$ is
the black hole mass.  This 
implies a unit of mass, length, and time of $M$, $G M/c^2$,
and $G M/c^3$, respectively.  The angular momentum $J$ of
the black hole is described by $a \equiv J c/G M^2$, where
$-1 < a < 1$.  It is convenient to define, like NT, the
following relativistic correction factors that become
unity in the nonrelativistic limit:
\begin{equation}
\sA \equiv 1 + a^2/r^2 + 2 a^2/r^3,
\end{equation}
\begin{equation}
\sC \equiv 1 - 3/r + 2 a/r^{3/2},
\end{equation}
\begin{equation}
\sD \equiv 1 - 2/r + a^2/r^2,
\end{equation}
and
\begin{equation}
\mu \equiv 1 + a^2 \cos^2\theta/r^2.
\end{equation}
In Boyer-Lindquist coordinates, the Kerr metric is
\begin{eqnarray}
ds^2 = & -(1 - {2\over{r \mu}}) dt^2 - {4 a \sin^2\theta
        \over{r \mu}} dt d\phi + {\mu\over{\sD}} dr^2
        + r^2 \mu d\theta^2 + \\ 
        & r^2 \sin^2\theta \left(
                1 + {a^2\over{r^2}} + {2 a^2 \sin^2\theta
                \over{r^3 \mu}}\right) d\phi^2.
\end{eqnarray}
The non-zero contravariant components are
\begin{equation}
\begin{array}{ccc}
g^{tt} = -1 - 2 (1 + a^2/r^2)/r\sD \mu, &
g^{t\phi} = -2 a/r^3\sD \mu, &
g^{rr} = \sD/ \mu, \\
g^{\theta\theta} = 1/r^2 \mu, &
g^{\phi\phi} = (1 - 2/r \mu)/r^2\sin^2\theta\sD.
\end{array}
\end{equation}
Often we shall only require the metric in the equatorial
plane, where 
\begin{equation}
ds^2 = -{\sD\over{\sA}} dt^2 + r^2\sA(d\phi - \omega d t)^2
        + {1\over{\sD}}dr^2,
\end{equation}
and
\begin{equation}
\omega \equiv {2 a \over{\sA r^3}}
\end{equation}
measures the rate of frame dragging by the hole.  
In expanded form,
\begin{equation}
ds^2 = -(1 - {2\over{r}}) dt^2 - {4 a\over{r}} dtd\phi
        + {1\over{\sD}} dr^2 + r^2 \sA d\phi^2.
\end{equation}
The nonzero contravariant components of the metric in
the equatorial plane are
\begin{equation}
\begin{array}{cccc}
g^{tt} = -\sA/\sD, & 
g^{t\phi} = -2 a/r^3\sD, &
g^{rr} = \sD, &
g^{\phi\phi} = (1 - 2/r)/r^2\sD.
\end{array}
\end{equation}
In the equatorial plane, the horizon lies 
at the outer root of $\sD = 0$, i.e. $r = 1 + \sqrt{1 - a^2}$.
The boundary of the ``ergosphere'', where the world lines of
observers at constant Boyer-Lindquist coordinates $(r,\theta,\phi)$ 
become spacelike (and therefore unphysical) lies at $r = 2$ in
the equatorial plane.

\subsection{Basic Flow Symmetry, Vertical Averaging}

Our goal is to accurately describe the flow of fluid in
the highly relativistic regime, close to the horizon.  To do
so we must make some simplifying assumptions.  First, we assume
the angular momentum of the accreting fluid is aligned with
the angular momentum of the black hole.
\footnote{
We make this assumption because the full non-aligned problem is
intractable.  Alignment of the flow due to gravitomagnetic 
precession and viscosity (the Bardeen-Petterson effect, see
\cite{bp75}, \cite{kp85}) is not likely to be important for
the hot flows considered here.
}
Second, we assume that the flow, described by its four-velocity
$(u^t,u^r,u^\theta,u^\phi)$ is in the mean axisymmetric and 
symmetric about the equatorial plane.  Then $u_\theta$ vanishes 
at the equatorial plane.

We also vertically average the flow, a standard approximation 
in accretion flows with angular momentum.  This is best explained 
by example.  Consider the particle number conservation
equation:
\begin{equation}
(\rho u^\mu)_{;\mu} = 0.
\end{equation}
Here $\rho$ is the ``rest-mass density''.  Then
\begin{equation}\label{PNCONSA}
(\rho u^\mu)_{;\mu} = {1\over{\sqrt{g}}}(\sqrt{g} u^\mu)_{,\mu}
	= {1\over{r^2}}(r^2 \rho u^r)_{,r} 
	+ {1\over{\mu\sin\theta}}(\mu\sin\theta \, u^\theta)_{,\theta} 
	= 0,
\end{equation}
where $g = |{\rm Det}(g_{\mu\nu})| = r^4 \sin\theta^2\mu^2$.

We now perform a ``vertical'' average by integrating over the volume 
between $r$ and $r + \delta r$.  Then the second term in equation 
(\ref{PNCONSA}) vanishes.  Now define $H_\theta$, the characteristic
angular scale of the flow about the equator, and assume this to be
the same for all flow variables $f$.  The vertical averaging approximation
consists in taking
\begin{equation}
\int d\theta d\phi \sqrt{g}\, f \simeq 4\pi H_\theta \, f(\theta = \pi/2),
\end{equation}
whence
\begin{equation}
(4 \pi H_\theta r^2 \rho u^r)_{,r} = 0.
\end{equation}
Integrating once in radius,
\begin{equation}
4\pi H_\theta r^2 \rho u^r = -\dot{M},
\end{equation}
where the constant $\dot{M}$ is the ``rest-mass accretion rate''.

\subsection{Frames}

We use four separate frames in our calculations.  The first 
and most important is the Boyer-Lindquist coordinate frame (BLF), 
where most direct calculations are done.  

Second is the locally nonrotating frame (LNRF), an 
orthonormal tetrad basis carried by observers who live at 
constant $\theta$ and $r$, but at $\phi = \omega t + const.$,
so they are dragged in azimuth by the black hole.  Their
world lines are always timelike, even in the ergosphere.
Explicit transformations between the LNRF and BLF are given 
by \cite{bpt72}.

A third frame is obtained by an azimuthal Lorentz boost from
the LNRF into another tetrad basis that corotates with the fluid
(the corotating frame, or CRF).  The CRF has velocity $\beta_\phi$ 
with respect to the LNRF; we also define $\gamma_\phi \equiv 
(1 - \beta_\phi^2)^{-1/2}$.

Finally we have the local rest frame (LRF) of the fluid, obtained
by a radial Lorentz boost from the CRF.  The LRF has radial 
velocity $V < 0$ with respect to the CRF, so we define 
$\beta_r \equiv V$ and $\gamma_r \equiv (1 - \beta_r^2)^{-1/2}$.

Two velocity variables are needed to describe the flow
in the equatorial plane.  We choose $V$ (as do ACGL), the radial 
velocity of the fluid measured in the CRF, and $l \equiv u_\phi$ (as
do PA), the angular momentum.

All the contravariant and covariant components of the velocity
field can now be expressed in terms of $l$ and $V$.  We
find, using $\gamma \equiv \gamma_\phi\gamma_r$,
\begin{eqnarray}\label{COVVEL}
\left(u_t,u_r,u_\theta,u_\phi \right) & = &
\left(
-\gamma \sqrt{\sD\over{\sA}} - l\omega, \quad
{V \over{\sqrt{\sD (1 - V^2)}}}, \quad
0, \quad
l \quad
\right), \\
\left(u^t, u^r, u^\theta, u^\phi \right) & = &
\left(
\gamma \sqrt{\sA\over{\sD}}, \quad
V \sqrt{\sD\over{1 - V^2}}, \quad
0, \quad
{l\over{r^2\sA}} + \omega\gamma\sqrt{\sA\over{\sD}} \quad
\right).
\end{eqnarray}
In terms of $l$ and $V$,
\begin{equation}
\gamma^2 = {1\over{1 - V^2}} + {l^2\over{r^2 \sA}},
\end{equation}
\begin{equation}
\beta_\phi = {l\over{r\sqrt{\sA}\gamma}}.
\end{equation}
Recall that $\sE \equiv -u_t$ is the ``energy at infinity'', and
that $l$ and $\sE$ are conserved along geodesics.

Finally, it is useful to have explicit expressions for the
LRF basis vectors.  The contravariant basis vectors are
\begin{equation}\label{CONBASA}
{e^\mu}_{(t)} = (\gamma\sqrt{{\sA\over{\sD}}}, \quad
	\beta_r\gamma_r \sqrt{\sD}, \quad
	0, \quad
	{2 a \gamma\over{r^3\sqrt{\sA\sD}}} +
	{\beta_\phi \gamma\over{r\sqrt{\sA}}})
\end{equation}
\begin{equation}\label{CONBASB}
{e^\mu}_{(r)} = (\beta_r\gamma\sqrt{{\sA\over{\sD}}}, \quad
	\gamma_r\sqrt{\sD},  \quad 0, \quad
	{2 a \beta_r\gamma\over{r^3\sqrt{\sA\sD}}} +
	{\beta_r\beta_\phi\gamma\over{r\sqrt{\sA}}})
\end{equation}
\begin{equation}\label{CONBASC}
{e^\mu}_{(\theta)} = (0, \quad 0,\quad {1\over{r}},\quad 0)
\end{equation}
\begin{equation}\label{CONBASD}
{e^\mu}_{(\phi)} = (\beta_\phi\gamma_\phi\sqrt{\sA\over{\sD}},\quad 
	0,\quad 0,\quad 
	{2 a \beta_\phi\gamma_\phi\over{r^3 \sqrt{\sA\sD}}} +
	{\gamma_\phi\over{r\sqrt{\sA}}}).
\end{equation}
The covariant basis vectors are
\begin{equation}\label{COVBASA}
{e_\mu}^{(t)} = (\gamma\sqrt{\sD\over{\sA}} +
	{2 a \beta_\phi\gamma\over{r^2\sqrt{\sA}}},\quad 
	-{\beta_r\gamma_r\over{\sqrt{\sD}}},\quad  0,\quad 
	-\beta_\phi\gamma r\sqrt{\sA})
\end{equation}
\begin{equation}\label{COVBASB}
{e_\mu}^{(r)} = (-\beta_r\gamma\sqrt{\sD\over{\sA}} -
	{2 a \beta_r \beta_\phi\gamma\over{r^2\sqrt{\sA}}},\quad 
	{\gamma_r\over{\sqrt{\sD}}},\quad  0,\quad 
	\beta_\phi\beta_r\gamma r\sqrt{\sA})
\end{equation}
\begin{equation}\label{COVBASC}
{e_\mu}^{(\theta)} = (0,\quad 0,\quad r,\quad 0)
\end{equation}
\begin{equation}\label{COVBASD}
{e_\mu}^{(\phi)} = (-\beta_\phi\gamma_\phi\sqrt{\sD\over{\sA}}
	- {2 a \gamma_\phi\over{r^2\sqrt{\sA}}},\quad  0,\quad  0,\quad 
	\gamma_\phi r \sqrt{\sA})
\end{equation}
Recall that the basis vectors allow one to transform back and forth
from the LRF via $v_{(a)} = e_{(a)}^\mu v_\mu$ and
$v_\mu = e^{(a)}_\mu v_{(a)}$.

\section{Basic Dynamical Equations}

The basic equations of relativistic, viscous hydrodynamics
are the energy-momentum conservation equations
\begin{equation}\label{USELESS}
{T^{\mu\nu}}_{;\nu} = 0,
\end{equation}
and the particle number conservation, or continuity, equation:
\begin{equation}
(\rho u^\mu)_{;\mu} = 0.
\end{equation}
We adopt the convention that $\rho$ is the ``rest mass density'' 
and $u$ is the internal energy per unit proper volume, so the 
total density of mass-energy is $\rho + u$.  We also define
the pressure $p$ and the relativistic enthalpy 
$\eta \equiv (\rho + u + p)/\rho$.  Then the stress-energy tensor is
\begin{equation}
T^{\mu\nu} = p g^{\mu\nu} + \rho\eta u^{\mu} u^{\nu} +
	t^{\mu\nu},
\end{equation}
where $t^{\mu\nu}$ is the ``viscous'' stress-energy tensor.  It
includes terms due to small-scale velocity and magnetic field
fluctuations.  We neglect contributions to the stress-energy
tensor from the energy flux and from large-scale electromagnetic
fields such as those found in MHD winds. 

In relativistic Navier-Stokes flow with zero bulk viscosity
\begin{equation}\label{NSSET}
t^{\mu\nu} = -2\lambda\sigma^{\mu\nu},
\end{equation}
where $\lambda$ is the ``coefficient of dynamic viscosity'',
\footnote{See \cite{is72} for a discussion of relativistic
microscopic viscosity.}
and 
\begin{equation}\label{SIGDEF}
\sigma_{\alpha\beta} = {1\over{2}}\left(
        u_{\alpha;\mu} h^\mu_\beta + u_{\beta;\mu} h^\mu_\alpha
        \right) - {1\over{3}}\Theta h_{\alpha\beta},
\end{equation}
is the shear tensor.  Here
\begin{equation}
\Theta \equiv {u^\alpha}_{;\alpha},
\end{equation}
and
\begin{equation}
h^{\alpha\beta} \equiv g^{\alpha\beta} 
	+ u^\alpha u^\beta
\end{equation}
is the projection tensor.  Our turbulent shear stress tensor does
not have this form, that is, it is not a relativistic Navier-Stokes 
viscous stress tensor.  Instead we write down a turbulent stress
tensor, described in \S 4, that is both the minimal departure from
the inviscid equations that allows angular momentum transport and
the simplest form that preserves causality.

\subsection{Particle Number Conservation}

As already discussed, the particle number conservation 
equation can be reduced to the form
\begin{equation}\label{PNCONS}
-4\pi r^2 \rho u^r H_\theta = \dot{M}.
\end{equation}
Expressing $u^r$ in terms of the preferred dependent 
variable $V$, 
\begin{equation}
4\pi r^2 \rho H_\theta V \left(\sD\over{1 - V^2}\right)^{1/2}
	= -\dot{M}.
\end{equation}

A closely related equation may be derived with the aid
of the Killing vector $\xi^\mu_t = (1,0,0,0)$.  Then
\begin{equation}
({T^\nu}_\mu \xi^\mu_t)_{;\nu} = 0 = 
{1\over{r^2}}(r^2 {T^r}_t)_{,r} 
+ {1\over{\mu\sin\theta}}(\mu\sin\theta \,{T^\theta}_t)_{,\theta}.
\end{equation}
Vertically averaging, the second term vanishes.
Then integrating once in $r$ and expanding we find
\begin{equation}\label{MEFLUX}
4\pi H_\theta r^2 [-(\rho + u + p)\sE u^r + {t_t}^r]
= \dot{E}.
\end{equation}
This equation expresses the constancy of mass-energy flux 
$\dot{E}$ with radius; $\dot{E}$ is the actual rate of change
of the black hole mass.  If the flow is cold and slow at 
large radius, $\dot{E} \approx \dot{M}$.  

Subtracting equation (\ref{PNCONS}) from equation (\ref{MEFLUX}) 
then dividing by equation (\ref{PNCONS}) gives the following
relation:
\begin{equation}\label{BERNOULLI}
-{{t_t}^r\over{\rho u^r}} + \sE - 1 + \left({u + p
\over{\rho}}\right)\sE = {\dot{E} - \dot{M}\over{\dot{M}}}.
\end{equation}
This reduces to the Bernoulli equation in the inviscid limit.

\subsection{Energy Equation}

The energy equation is the component of equation (\ref{USELESS})
parallel to the fluid four-velocity:
\begin{equation}
u_\mu {T^{\mu\nu}}_{;\nu} = 0.
\end{equation}
Using standard manipulations (e.g. \cite{ell71}), this
can be reduced to 
\begin{equation}
u^r {d u\over{d r}} - u^r {(u + p)\over{\rho}} {d\rho\over{d r}}
	= \Phi - \Lambda
\end{equation}
where, provided that $t^{\mu\nu}$ is trace-free, as we shall 
assume,
\begin{equation}\label{DISSDEF}
\Phi = -t^{\mu\nu} \sigma_{\mu\nu}
\end{equation}
is the dissipation function and $\Lambda$ is the cooling function.

Two variables are required to represent the
thermodynamic state of the fluid.  We choose $\rho$, the rest-mass
density, and the scaled temperature $T$.  In dimensional form, $T$
is related to the ordinary kinetic temperature $T_K$ by $T \equiv 
k T_K/\bar{m} c^2$, where $k$ is Boltzmann's constant and $\bar{m}$ 
is the mean molecular weight.  

The pressure is given by the ideal gas equation of state 
$p = \rho T$.  We fold the magnetic contribution to the
pressure into $p$ and any magnetic contribution to the energy
per unit proper volume into the internal energy $u$.
We must now specify $u(\rho,T)$.  We require an
equation of state that is polytropic in the nonrelativistic limit 
with adiabatic index $\gamma_o$.  Since the magnetic energy
is included in the internal energy, $\gamma_o$ may differ from
$5/3$.  Thus we set
\begin{equation}
u = \rho T g(T) \equiv \rho T \left({4/(\gamma_0 - 1) + 15 T\over{
	4 + 5 T}}\right).
\end{equation}
This formula fits the exact relativistic equation of state for an
ideal relativistic Boltzmann gas (see, e.g., \cite{ser86}), for 
which $\gamma_o = 5/3$, to everywhere better than $2\%$. 
\footnote{
An even better fit to the ideal relativistic Boltzmann gas equation
of state is $g(T) = (12 + 45 T + 45 T^2)/(8 + 20 T + 15 T^2)$, which 
has a maximum relative error of $7 \times 10^{-4}$.
}
The sound speed is then
\begin{equation}
c_s^2 = {\Gamma p\over{\eta \rho}} = {\Gamma T\over{\eta}},
\end{equation}
where
\begin{equation}
\Gamma \equiv \left. {\del\ln p\over{\del \ln\rho}}\right|_{s} = 
	1 + {1\over{g + T (d g/d T)}}.
\end{equation}
When $\gamma_o = 5/3$,
our $\Gamma$ agrees with Service's to within $0.5\%$, and our
sound speed agrees with hers to within $1.3\%$.

The energy equation can now be reduced to 
\begin{equation}\label{ENCONS}
V \left(\sD\over{1 - V^2}\right)^{1/2}
\left({\del u\over{\del T}} {d T\over{d r}} - 
{p\over{\rho}}{d\rho\over{d r}}\right) = \Phi - \Lambda.
\end{equation}
The dissipation function can be evaluated once we specify
the viscous stress.  

Like NKH, we do not include cooling explicitly, but set
$\Phi - \Lambda = f\Phi$, where $f \le 1$ is a 
constant factor that we vary to estimate the effect
of cooling.  Our energy equation is essentially the same
as ACGL, except that we include the relativistic equation
of state, and our dissipation function is different.
PA do not solve an energy equation; instead they use a
polytropic equation of state.

\subsection{Radial Momentum}

The radial momentum equation is the radial component of the
projection of equation (\ref{USELESS}) into the space normal
to the four-velocity:
\begin{equation}
h_{r\mu} (T^{\mu\nu})_{;\nu} = 0.
\end{equation}
Using standard manipulations (we must evaluate some connection
coefficients), and neglecting the viscous acceleration, 
the radial momentum equation becomes
\begin{equation}\label{RMCONS}
{V\over{1 - V^2}} {d V\over{d r}} = f_r - {1\over{\rho\eta}}
	{d p\over{d r}},
\end{equation}
where
\begin{equation}
f_r \equiv -{1\over{r^2}}{\sA \gamma_\phi^2\over{\sD}}
	(1 - {\Omega\over{\Omega}}_+) (1 - {\Omega\over{\Omega}}_-)
\end{equation}
and $\Omega \equiv u^\phi/u^t = \omega + 
l \sD^{1/2}/r^2\sA^{3/2}\gamma$, and $\Omega_\pm =
\pm (r^{3/2} \pm a)^{-1}$ is the rotation frequency, as 
observed at large radius, of circular, planar orbits.

Our radial momentum equation obviously reduces correctly to the
nonrelativistic limit.  It agrees with PA's radial momentum
equation, although theirs is expressed in different variables.
It also agrees with ACGL's, although ACGL take $\eta = 1$.

\subsection{Angular Momentum}

An angular momentum equation can be derived with the aid of
the azimuthal Killing vector $\xi^\mu_\phi = (0,0,0,1)$.  
We have
\begin{equation}
({T^\nu}_\mu \xi^\mu_\phi)_{;\nu} = 0 =
({T^\nu}_\phi)_{;\nu} = 
{1\over{\sqrt{g}}} (\sqrt{g}{T^\nu}_\phi)_{,\nu} =
{1\over{r^2}}(r^2 {T^r}_\phi)_{,r} 
+ {1\over{\mu\sin\theta}}(\mu\sin\theta \,{T^\theta}_\phi)_{,\theta}.
\end{equation}
Vertically averaging, the second term vanishes, and integrating
once and using the definition of the stress tensor, we obtain
\begin{equation}\label{AMCONS}
\dot{M} l\eta - 4\pi H_\theta r^2 {t^r}_\phi = \dot{M} j.
\end{equation}
Here $\dot{M} j$ is the total inward flux of angular momentum;
$j$ is treated as an eigenvalue of the problem and must be
self-consistently obtained.

Our angular momentum equation agrees with PA and ACGL, except
that ACGL assume $\eta = 1$.  

\subsection{Vertical Structure}

We assume vertical equilibrium, and use the expression for
the simplified equilibrium scale height derived by ALP 
under the assumption that $u^\theta$ and ${u^\theta}_{,
\theta}$ are small.
A convenient way to to express ALP's result is to define
an effective vertical frequency via
\begin{equation}\label{VERT}
H_\theta^2 = {p\over{\rho\eta r^2 \nu_z^2}}.
\end{equation}
Then ALP find
\footnote{
This expression is closely related to a well-known
integral of the motion for geodesics in the Kerr geometry:
$Q = u_\theta^2 + \cos^2\theta \left({l^2\over{\sin^2\theta}} 
- a^2 (\sE^2 - 1)\right)$ [\cite{car68}].  In the limit of small
$\delta\theta = \theta - \pi/2$, $Q \simeq u_\theta^2 + r^4 
\delta\theta^2 \nu_z^2$.  Evidently $Q$ is proportional to a 
vertical energy of oscillation.  
}
\begin{equation}
\nu_z^2 = {l^2 - a^2 (\sE^2 - 1)\over{r^4}}.
\end{equation}
In the limit of a relativistic, thin disk this reduces to
$\nu_z^2 = r^{-3} (1 - 4 a/r^{3/2} + 3 a^2/r^2)/\sC$, which
agrees with the result of \cite{rh96}, and obviously reduces
correctly to the nonrelativistic limit.

ALP's result for the equilibrium scale height differs from PA's.
The reason is that ALP neglect $u^\theta$ and its derivatives,
while PA work in a quasi-cylindrical coordinate system 
$(t,r,z,\phi)$ and thus neglect $u^z$ and its derivatives.  
The two assumptions are not equivalent.  If the flow is 
quasi-spherical, then ALP's approach is superior.

The major defect of all these approaches to vertical structure
is that they assume vertical equilibrium.  This assumption
is not justified close to the event horizon.  Inflow there is 
so rapid that inertial terms become important, and the scale 
height effectively ``freezes out.'' ALP provide an expression 
for the evolution of the scale height that takes this effect 
into account, but we have not used this because it introduces 
significant mathematical complications.

\section{Turbulent Shear Stress}

The key to calculating the structure of steady,
rotating accretion flows lies in understanding the nature of
the viscous stress ${t^r}_\phi$ that allows the angular 
momentum of the flow to evolve.  In our view the viscous stress 
is likely caused by MHD turbulence initiated by the Balbus-Hawley
instability (\cite{bh91}).  In this case, the ``viscous'' stress is 
really a time-averaged turbulent Reynolds and Maxwell stress
(see \cite{bgh94}).

The viscous stress is commonly assumed to have the Navier-Stokes 
form (see eq. [\ref{NSSET}]).  One difficulty with this assumption
is that, as pointed out by \cite{nar92}, the resulting system
of equations are acausal: they propagate information at infinite 
speed.  This has the undesirable consequence that one must apply 
a boundary condition at the event horizon (e.g. \cite{nkh}).  

Various workers have sought to remedy this defect, including
\cite{nar92}, \cite{nlk}, \cite{ps94} (hereafter PS),
\cite{ki94}, \cite{kat94}, and Narayan (1996, private
communication).  We follow PS's line of reasoning 
because it is straightforward to generalize it to the 
relativistic case.  

\subsection{Causal Viscosity: Nonrelativistic Case}

PS introduce a phenomenological equation for the evolution of 
the viscous stress that allows it to relax toward an equilibrium 
value.  Let us review 
their result in our notation, temporarily assuming that the
velocities are nonrelativistic and the geometry is Euclidean.
Then PS's equation for the viscous stress $S = \ss$ is
\begin{equation}\label{NRSTEVOLVA}
{D S\over{D\tau}} = -{S - S_o\over{\tau_r}},
\end{equation}
where $D/D\tau$ is the convective derivative, $S_o$ is the 
equilibrium value of the stress and $\tau_r$ is the relaxation 
timescale.  In a steady state
\begin{equation}\label{NRSTEVOLV}
V {d S\over{d r}} = -{S - S_o\over{\tau_r}}.
\end{equation}
Now take $d/dr$ of the angular momentum equation to obtain
\begin{equation}\label{DIFFAM}
{\dot{M}\over{2\pi}}{d l\over{d r}} = 2 H_\theta\left(
	2 r S + r^2 {d S\over{d r}}\right),
\end{equation}
neglecting terms of order $d H_\theta/d r$.
Solve for $d S/dr$ in equation(\ref{DIFFAM}),
eliminate $d S/dr$ from equation(\ref{NRSTEVOLV}), and
use $S_o = -\rho\nu r^2 d\Omega/dr$ to find
\begin{equation}\label{NRCAUSST}
S = {\rho\nu r \over{1 - 2\tau V/r}}\left(
	-r {d\Omega\over{d r}} + {V^2\over{c_\nu^2}}
	{1\over{r}}{d l\over{d r}}\right),
\end{equation}
where $c_\nu = \sqrt{\nu/\tau_r}$ is the speed at which
viscous effects propagate (see the discussion of PS).  
If $\tau_r \simeq \Omega^{-1}$
(as one might expect for Balbus-Hawley induced turbulence), 
and $\nu \simeq \alpha c_s^2/\Omega$, then $c_\nu \simeq 
c_s\alpha^{1/2}$.  Rewriting equation(\ref{NRCAUSST})
slightly,
\begin{equation}
S ={\rho\nu r \over{1 - 2\tau V/r}}\left[ -{1\over{r}}{d l\over{d r}}
	\left(1 - {V^2\over{c_\nu^2}}\right) +
	{2 l\over{r^2}}\right].
\end{equation}
Evidently this viscosity prescription introduces a singular 
point into the angular momentum equation at $V^2 = c_\nu^2$,
since the coefficient of $d l/d r$ in the angular momentum
equation vanishes there.
This requires a new boundary condition at the singular point.  
The new boundary condition replaces the old, objectionable 
boundary condition at the surface of the accreting object.

\subsection{Causal Viscosity: Relativistic Case}

We now generalize PS's causal stress prescription to the
relativistic case.  Begin with the viscous stress-energy tensor 
measured in the LRF, $t_{(a)(b)}$.  We assume that all 
components vanish except $t_{(r)(\phi)} = t_{(\phi)(r)}$.
This is the minimal modification of the
inviscid equations that produces angular momentum
transport.  Notice that the LRF is the only safe place to
make such modifications, since an arbitrary modification
of the stress tensor in another frame can have unintended
side effects.  Then, with the 
aid of equations (\ref{COVBASB}) and (\ref{COVBASD}), we find 
\footnote{
We disagree with \cite{ap90}, who argue that the viscous 
stress (indeed, the viscosity) vanishes at the horizon.  
Measured in the LRF, the viscous stress will not generally vanish 
at the horizon because the horizon is not in any way
a special location from the point of view of the fluid,
which falls freely through it.  Transforming
to the BLF, one finds that for general $t_{(a)(b)}$,
$\ss = \gamma r \sqrt{\sA\sD}(\css + \beta_r\beta_\phi
\gamma_r t_{(r)(r)})$.  For geodesics $\gamma \sim
\gamma_r \sim \sD^{-1/2}$ near the horizon, so if $\css$ 
and $t_{(r)(r)}$ are finite at the horizon, so is $\ss$.
}
\begin{equation}
\ss = \gamma r \sqrt{\sA \sD} \css \equiv F \css.
\end{equation}

We now identify $\css$ with $S$, the component of the
stress that obeys the relativistic analogue of 
equation (\ref{NRSTEVOLVA}):
\begin{equation}
{D S\over{D \tau}} = -{S - S_o\over{\tau_r}},
\end{equation}
where now $D/D\tau = u^\mu ()_{;\mu}$.  Then in a
steady state
\begin{equation}\label{STEVOLV}
u^r {d S\over{d r}} = -{S - S_o\over{\tau_r}}.
\end{equation}
The rest of the argument proceeds as before: we 
differentiate the angular momentum equation, solve for
$d S/d r$, eliminate $d S/d r$ from equation 
(\ref{STEVOLV}), and solve for $S$.  We find
\begin{equation}\label{SINIT}
S = {1\over{1 - \tau_r u^r (2/r + d\ln F/dr)}}\left(
S_o + {\rho \tau_r (u^r)^2\over{F}}
	{d (l\eta)\over{d r}}\right).
\end{equation}
We must now specify $S_o$.

The most natural form for the equilibrium value of the
LRF shear stress is the Navier-Stokes value:
\begin{equation}\label{NSSTRESS}
S_o = -2\rho\eta\nu \sigma_{(r)(\phi)},
\end{equation}
where $\nu = \alpha c_s r H_\theta$, following the usual 
$\alpha$ prescription.  The calculation of $\sigma_{(r)(\phi)}$ 
is a lengthy, but important, matter that is left to the Appendix.  
The result is also complicated and left in the Appendix.
Here we abbreviate it by
\begin{equation}
\sigma_{(r)(\phi)} \equiv \sigma \equiv
	\sigma_N + \sigma_L {d l\over{d r}}
	+ \sigma_V {d V\over{d r}}.
\end{equation}
In the nonrelativistic limit, $\sigma = (1/2) r d\Omega/d r$,
while in the thin disk limit $\sigma = -(3/4) r^{-3/2}
\sD/\sC$ (\cite{nt73}).
\footnote{
One can show that $-\sigma$ is the maximum growth rate
for the Balbus-Hawley instability in a thin Kerr disk.
The factor $\sD/\sC$ varies from $1$ at large radius to
$4/3$ at the last stable orbit.  In the mechanical analogy
for the Balbus-Hawley instability developed by \cite{bh92}
this result is unaffected by radial motions, to first order
in the eccentricity.
}

Now we recover the full expression for the shear stress
$S$.  Substituting in equation (\ref{SINIT}) for $S_o$,
we find
\begin{equation}\label{STRESS}
S = {\rho\eta\nu\over{1 - \tau_r u^r (2/r + d\ln F/dr)}}
\left\{ {d l\over{d r}}\left(
	-2\sigma_L + {(u^r)^2\over{c_\nu^2}}{1\over{F}}
	\right) - 2 \sigma_N - 2 \sigma_V {d V\over{d r}}
	+ {(u^r)^2\over{c_\nu^2 F}}{d\ln\eta\over{d r}}
	\right\}.
\end{equation}
The coefficient of $d l/d r$ vanishes, and so there is
a singular point in the flow, where $(u^r)^2 \approx c_\nu^2$.

\subsection{Dissipation Function}

The dissipation function $\Phi$ is best calculated in
the LRF, where
\begin{equation}
\Phi  = -t^{(a)(b)} \sigma_{(a)(b)}
\end{equation}
Since most of the components of $t$ vanish in the LRF,
and the metric is Minkowski there, one finds
\begin{eqnarray}\label{DISSFUNC}
\Phi & = & -2 S \sigma \nonumber \\
	& = & \rho \eta \nu \left(
		4 \sigma^2 - 2 \sigma {(u^r)^2\over{
		c_\nu^2 \eta f}}{d (l\eta)\over{d r}}\right)
\end{eqnarray}
If $\sigma < 0$ and $d (l\eta)/dr > 0$ then $\Phi$ is 
positive.  

It turns out that the dissipation function is not always
positive in our solutions.  In particular, $\sigma$ sometimes
changes sign close to the horizon, an effect first noticed
by \cite{al88}.  Since the stress is not proportional to
the shear rate, it need not change sign at the same point.
The dissipation function then goes negative, indicating that
the stress is transporting angular momentum outwards against
the shear.  In part this reflects the inadequacy of our
phenomenological equation for the evolution of the viscous 
stress.  In part this may be a physical effect, however,
in the sense that small scale magnetic fields embedded in
the flow are ``wound up'' by the shear and briefly unwind,
reducing the internal energy of the fluid (into which we
folded the magnetic energy).  But it seems unlikely that this
unwinding is fully reversible.  As we shall see in the 
sample numerical solutions, this effect has no practical
consequences since it only occurs close to the horizon,
where the inflow time is short compared to the heating
time and so the flow is nearly adiabatic.

\section{Critical Points}

The accretion flow must pass through two critical points.
The first is the usual sonic point, which is made rather
complicated by the presence of relativistic terms.  The second is
the ``viscous point'' that originates in the vanishing
of the coefficient of $d l/d r$ in the angular momentum
equation.  Here we develop explicit expressions for the
associated boundary conditions.

Before proceeding, it is convenient to make two more
approximations.  These approximations eliminate terms
that weakly couple the angular momentum equation to the
other equations and enormously complicate the evaluation
of the critical points.  First, we replace $d\ln F/dr$ 
in equation (\ref{STRESS}) by $1/(r\gamma\sqrt{\sD})$.  
This approximation is good to about $15\%$ inside $r = 10$ 
and far better outside.  Second, we replace
$d\ln\eta/dr$ in equation (\ref{STRESS}) by $-(\eta-1)/
\eta r$.  This approximation is good to about a factor of
two.  These terms are important mainly in the relativistic
regime, where the shear stress is dynamically unimportant
and the flow is nearly adiabatic.

\subsection{Sonic Point}

In order to obtain the sonic point conditions, we need to find an
expression for the radial velocity gradient which involves no other
derivatives.  Therefore, we must find an explicit expression for the
pressure derivative in (\ref{RMCONS}).  To accomplish this, we first
gather together the conservation equations for particle number and
energy (\ref{PNCONS}) and (\ref{ENCONS}), and the vertical equilibrium
equation (\ref{VERT}), and write them in differential form:
\begin{equation}\label{DPNCONS}
{d \ln \rho \over d r} + {d \ln H_\theta \over d r} + {2 \over r} +
{1 \over 1 - V^2} {d \ln V \over d r} + {1 - a^2/r \over r^2 \sD} = 0
\end{equation}
\begin{equation}\label{DENCONS}
(h(T) - 1) {d \ln T \over d r} - {d \ln \rho \over d r} = {f \Phi
\over \rho u^r T}
\end{equation}
\begin{equation}\label{DVERT}
2 {d \ln H_\theta \over d r} = (1 - T h(T) / \eta) {d \ln T \over d r}
- {2 \over r} - {d \ln \nu_z^2 \over d r}.
\end{equation}
We have used $h(T) \equiv 1 + g(T) + T dg/dT$, so that
$d \eta / dr = h(T) dT/dr$, and $(1/\rho) \del u/\del T = h(T) - 1$.
Notice that we have retained a term proportional to $d H_\theta/d r$;
this term is assumed small, but we must retain it for consistency
if we take equation (\ref{PNCONS}) as our basic particle number
conservation equation.

Using (\ref{DVERT}) to eliminate $d \ln H_\theta / dr$ from
(\ref{PNCONS}), we can solve for $d \ln \rho / dr$ and $d \ln T /
dr$, and obtain $d \ln P / d r = d \ln \rho / d r + d \ln T / d
r$.  Substituting the result into (\ref{RMCONS}), we obtain
\begin{equation}
{V^2 \over 1 - V^2} {d \ln V \over d r} = f_r - {T \over \eta}
\left[f_1(T) \left( {1 \over 2}{d \ln \nu_z^2 \over d r} 
- {1 \over 1-V^2}{d \ln V \over d r} - {r - 1 \over r^2 \sD} \right) 
+ f_2(T) {f \Phi \over \rho u^r T} \right],
\end{equation}
where 
\begin{equation}
f_1(T) \equiv {h(T) \over h(T) - 1 + (1/2)(1 - T h(T) / \eta)}, \quad
f_2(T) \equiv {1 - (1/2)(1 - T h(T) / \eta) \over h(T) - 1 + (1/2)(1 - T
h(T) / \eta)}.
\end{equation}
In the nonrelativistic ($T \rightarrow 0$) limit, $f_1 = 2\gamma_0/
(\gamma_0 + 1)$ and $f_2 = (\gamma_0 - 1)/(\gamma_0 + 1)$.

Finally, we must move all terms proportional to $d \ln V / dr$ from
the right-hand to the left-hand side of this equation.  This is more
involved that it first appears because both $\Phi$ and $d \ln \nu_z^2
/ dr$ include terms proportional to $d \ln V / dr$.  First, $\Phi$ 
can be simplified by substituting the shear stress $S$ from 
(\ref{AMCONS}) into (\ref{DISSFUNC}), which allows us to write 
\begin{equation}
{f \Phi \over \rho u^r T} = {2 \sigma f (l \eta - j) \over F T}.
\end{equation}
We then divide $\sigma$ into a part which is proportional to
$dV / dr$ and a part which is not:
\begin{equation}
\sigma = \left[\sigma_V + \sigma_L \left({dl \over dr}\right)_V
\right] {dV \over dr} + \sigma_N + \sigma_L \left({dl \over
dr}\right)_N,
\end{equation}
where we have taken $dl/dr = (dl/dr)_N + (dl/dr)_V dV/dr$.  The terms
$(dl/dr)_N$ and $(dl/dr)_V$ are derived from (\ref{AMCONS}) using the
shear stress (\ref{STRESS}). Thus, the $d \ln V /
dr$ part of the $\Phi$ term is 
\begin{equation}
-{f_2(T) \over \eta} \left(\sigma_V + \sigma_L \left({dl \over
dr}\right)_V\right) {2 f (l \eta - j) V \over F}{d \ln V \over dr}.
\end{equation}
Following a similar procedure, we find that the $d \ln \nu_z^2 /dr$
term has a $d \ln V /dr$ part
\begin{equation}
-{T f_1(T) \over 2 \eta} \left({d \ln \nu_z^2 \over dr}\right)_V {dV
\over dr} = -{T f_1(T) V \over \eta \nu_z^2} \left[{l - \sE a^2
\Omega \over r^4} \left({dl \over dr}\right)_V - {\sE a^2 \over
r^4} \left({d \sE \over dr}\right)_V \right] {d \ln V \over dr}.
\end{equation}

Solving for $d \ln V /dr$, we obtain
\begin{equation}
{d \ln V \over dr} = {N_s\over{D_s}},
\end{equation}
where
\begin{equation}\label{SONNUM}
N_s  = f_r - {T f_1(T) \over \eta}
\left[\left( {1 \over 2}\left({d \ln \nu_z^2 \over d r}\right)_N
- {r - 1 \over r^2 \sD} \right) \right]
- {f_2(T) \over \eta} \left(\sigma_N + \sigma_L
\left({dl \over dr}\right)_N\right) {2 f (l \eta - j) \over F},
\end{equation}
and
\begin{equation}\label{SONDEN}
D_s = {V^2 \over 1-V^2} + {T f_1(T) \over \eta}
\left({V \over 2} \left({d \ln \nu_z^2 \over dr}\right)_V - {1 \over
1-V^2} \right) + {f_2(T) \over \eta} \left(\sigma_V + \sigma_L
\left({dl \over dr}\right)_V\right) {2 f (l \eta - j) V \over F}.
\end{equation}
At the sonic point $N_s = D_s = 0$.  In the nonrelativistic limit 
the condition $D_s = 0$ reduces to
$V^2 = c_s^2 2\gamma_0/(\gamma_0 + 1)$, in agreement with NKH.

\subsection{Viscous Point}

The viscous point conditions are obtained by solving the angular
momentum equation (\ref{AMCONS}) for $dl/dr$.  We obtain
\begin{equation}
{d l\over{d r}} = {N_v\over{D_v}}
\end{equation}
where
\begin{equation}\label{VISCNUM}
N_v = {[1 - \tau_r u^r (2/r + d\ln F/dr)]
V \sD^{1/2} \over F \eta \nu (1-V^2)^{1/2}} 
+ 2 \sigma_N + 2 \sigma_V {dV \over dR} 
- {(u^r)^2 \over c_\nu^2 F} {d \ln \eta \over dr}.
\end{equation}
and
\begin{equation}
D_v = -2 \sigma_L + {(u^r)^2 \over c_\nu^2 F} 
\end{equation}
At the viscous point $N_v = D_v = 0$.  The conditions $D_v = 0$ is
equivalent to
\begin{equation}\label{VISCDEN}
{V^2 \over c_\nu^2} = \gamma (\sA/\sD)^{1/2} (1-V^2) (\sE - l \Omega).
\end{equation}
We take $c_\nu^2 = \alpha c_s^2$, and since $\alpha < 1$ the
viscous point will generally be located well outside the sonic
point.

\section{Example Numerical Solution}

In this section we find a typical numerical solution and use it to
study the importance of various relativistic terms, to check our
approximations, and evaluate the robustness of the input physics.  More
precisely, we have solved the system of equations (\ref{MEFLUX}),
(\ref{ENCONS}), (\ref{RMCONS}) and (\ref{AMCONS}) numerically, using a
relaxation scheme (see \cite{PG} for more details).  We have set
$f = 1$ (zero cooling), $\alpha = 0.1$, and the rotation 
parameter of the black hole $a = 0$.

The system is subject to five boundary conditions that allow us to
solve the first order differential equations for $V,l,\rho,T$ and for
the eigenvalue $j$ over a domain that extends from just outside the
event horizon to $r = 2\times 10^4 G M/c^2$.  Two boundary conditions
are provided by equations (\ref{SONNUM}) and (\ref{VISCNUM}), which
ensure the solution passes smoothly through the sonic and viscous
points, respectively.  Two additional boundary conditions fix 
$l$ and $T$ at the outer edge of the solution ($V$ must be determined
self-consistently there).  We set $T_{out} = 1.67 \times 10^{-5}$ and
$l_{out} = 57.75$.  These values are taken from the self-similar
solution of \cite{ny94}.  NKH have shown that global solutions with,
e.g., thin disk outer boundary conditions, rapidly approach this
self-similar solution, which then spans many decades in radius.  
Consistent with this, we find that the character of the solution at 
$r \lesssim 10^2$ is insensitive to the outer boundary conditions.   
A final boundary condition is obtained by normalizing the density so 
that the rest-mass accretion rate $\dot{M} = 1$.

The run of the basic variables $V,l,\rho,T$ with radius is shown in
Figure 1.
\footnote{Physical units may be recovered as follows: radial
velocity is $V c$, angular momentum is $l G M/c$, density is 
$\rho \dot{M} G/c^3$ (since the mass accretion rate is normalized
to 1), and temperature is $T \bar{m} c^2/k$.}
The upper left panel shows $V$, and the arrows indicate
the location of the sonic point at $r_s = 6.41$ (slightly outside
the last stable circular orbit at $r = 6$) and the viscous point at 
$r_v = 28.2$.  At the viscous point $V \approx 0.3 c_s$ and so it
lies far outside the sonic point.  The solution is, however, 
insensitive to the precise definition of $c_\nu$.  For example,
setting $c_\nu = 0.5 c_s$ changes $j$ by less than $4\%$.
The solution reaches $V = 1$ at the event horizon, which is located 
at $r = 2$.

The upper right panel in Figure 1 shows $l$ (heavy solid line),
$l\eta$ (dashed line), and the eigenvalue $j = 2.62$ (dotted line).
Recall that $l\eta$, rather than $l$, is conserved in the absence of
viscous torques, and that $\dot{M} j$ is the total inward flux of
angular momentum.  The difference $l\eta - j$ is proportional to the
shear stress; in this case the difference is a small positive number
at the horizon.  The lower left panel shows density, which very nearly
follows a power law $r^{-3/2}$ over the whole solution.  The lower
right panel shows the temperature, which reaches a maximum of $0.067$,
corresponding to a sound speed $c_s = 0.28 c$.  As we shall see, the
maximum temperature and density of the solution change significantly
with black hole rotation.

The run of some important dynamical quantities is shown in Figure 2.
The upper left panel shows the absolute value of the shear rate in the
LRF.  The shear changes sign at $r \approx 3$.  Also shown is the thin
disk approximation to the shear rate, equation (\ref{TDSHEAR}), which
lies too close to the actual shear rate to distinguish it on this
plot.  The maximum difference between the two is about one part in
$10^3$.  The upper right panel shows the causal shear stress as a solid
line and the equilibrium shear stress $S_0$ (proportional to the shear
rate, see equation [\ref{NSSTRESS}]) as a dashed line.  The causal
stress is relaxing toward $S_0$ but the fluid is also flowing inwards
toward regions of larger shear, so the causal stress always lies
slightly below $S_0$.  The lower left panel shows $H_\theta$.  The vertical
averaging approximation is accurate to of order $d\ln H_\theta/d\ln R$, which
is largest at the horizon.  The lower right panel shows $\Omega \equiv
u^\phi/u^t$.  For $a = 0$ we must have $\Omega = 0$ at the horizon, so
$\Omega$ goes through a maximum at $r \approx 3$.  This is consistent
with our statement that the shear rate in the LRF is very nearly
proportional to $d\Omega/d r$.

Thermodynamic quantities are shown in Figure 3.  These are the
relativistic enthalpy $\eta$, the pressure $p$, the entropy $S$ (minus
its value at the outer edge), and the dissipation function $\Phi$.  As
discussed earlier, the dissipation function goes negative at $r \approx
3.0$, where the shear changes sign.  This is an inevitable result of
our causal viscosity prescription.  It has little practical
significance, however, since, as can be seen from Figure 3, the flow is
nearly isentropic at $r < 3.0$.  This is because the radial inflow
time becomes short compared to the heating time.

This work contains two new physical ingredients as compared to NKH: a
causal viscosity prescription, and a full treatment of relativistic
effects.  We can evaluate the influence of each by solving a
simplified version of our equations which retains our causal viscosity
formulation, but eliminates all relativistic corrections and terms and
adopts the Paczy\'nski potential used by NKH.  We can compare the NKH
solution and the causal solution to find the effects of the causal
viscosity, and we can compare the causal solution and the full,
relativistic solution to find the effect of properly including
relativity.

The results are displayed in Figure 4, which shows the basic variables
$V,l,\rho,T$ in each of the three solutions:  the fully relativistic
solution (heavy solid line), the causal solution (long dashed line) and
the NKH solution (dotted line).  Also shown in the angular momentum
plot is a short dashed line, which is the angular momentum of a thin
disk (inside $r = 6$ the angular momentum is that of the last stable
orbit).  Before continuing with the comparison, notice that all the
ADAF solutions lie below the thin disk solution in angular momentum--
evidently they are all ``sub-Keplerian''.

Remarkably, the relativistic solution, causal solution, and NKH
solution are very similar at $r \gtrsim 10$.  This suggests that
the solutions are robust in the sense that significant changes in the
input physics cause only small changes in the solution.  One can
also check the robustness of the shear stress model by changing
the definition of $c_\nu^2$.  If we set $c_\nu^2 = c_s^2$ (rather
than $\alpha c_s^2$) we find that the values of the basic dependent
variables change by no more than $5.9\%$ at the inner boundary.
Again, this suggests that the solution is robust.

What is the
effect of the causal viscosity?  The comparison of the causal solution
and the NKH equations is not perfect because NKH take $\nu_z^2 =
\Omega_K^2$ (which diverges at $r = 2$), while we take $\nu_z^2 =
\Omega^2$.  The outcome of combining these two changes is to lower the
angular momentum slightly and decrease the radial velocity (therefore
increasing the density) in comparison to NKH.  What is the effect of
properly including the relativity?  The most obvious effect is in the
radial velocity, which diverges at $r = 2$ in the causal solution
because the potential gradient diverges there.  This causes the
density and temperature to go through a maximum and then decline to
zero at $r = 2$.  In the relativistic solution, by contrast, the
density and temperature increase monotonically in to the horizon.  The
next most obvious effect is in the angular momentum profile, which is
significantly lower at the horizon in the relativistic solution than
in the causal solution.  This is due to our retention of relativistic
enthalpy terms in the angular momentum equation; the angular momentum
per unit rest mass is $l\eta$ rather than $l$.

The similarity of all three ADAF solutions in Figure 4 should not lead
one to conclude that relativistic effects are not important.  Figure 5
shows the effect of varying $a$ from $-0.999$ to $0$ to $0.999$.  The
temperature at the horizon increases from $0.067$ when $a = 0$ to
$0.363$ when $a = 0.999$, while the density increases from $0.11$ at $a
= 0$ to $1.29$ at $a = 0.999$.  These changes suggest the possibility
that ADAF emission properties may depend strongly on black hole
rotation.  We defer a more detailed discussion of the effect of black
hole rotation to \cite{PG}.

We can now check our approximations for self-consistency.  Vertical
averaging cannot be checked directly, but the smallness of $|d\ln
H_\theta/d\ln r|$ compared to $1$ is a rough test of its validity
(this also tests the validity of our vertical equilibrium
approximation).  The maximum in $|d\ln H_\theta/d\ln r|$ is $0.697$ at
the horizon and much smaller outside (see Figure 2).  The situation
would be improved if the scale height were allowed to evolve
dynamically, so that $H_\theta$ could ``freeze out'' near the horizon.

We have also dropped viscous accelerations in the radial momentum
equation.  One would expect that the viscous term is of order $\alpha$
compared to the dominant terms, and a direct check confirms this.  The
viscous term generally produces an outward directed acceleration,
although in the inner parts of those solutions where the shear changes
sign (generally when $a \lesssim 0.7$) it produces an inward
acceleration.

We made two additional approximations prior to deriving the sonic
point conditions: (1) we set $d\ln F/dr \simeq 1/(r\gamma\sqrt{\sD})$;
(2) $d\ln\eta/dr \simeq -(\eta-1)/ \eta r$.  We have checked 
these approximations by multiplying the approximated term by $0.5$ 
and examining the effect on the solution.  For approximation (1)
we find the largest fractional change is $2.7\%$ in $T$ at the 
innermost radius.  For (2) we find the largest fractional change 
is $0.2\%$ in $l$ at the innermost radius.  This shows that the
approximations, whose accuracy was discussed at the beginning of
\S 5, do not affect the solution very much.

Finally, how important are the relativistic thermodynamic terms ($\eta$
and $g$) that we have worked so hard to include?  The $\eta$ term turns
out to be quite important, particularly for holes with rapid prograde
rotation.  When $a = 0.999$, the maximum value of $\eta$ is $2.26$;
this implies substantial corrections in the angular momentum and radial
momentum equations.  The relativistic equation of state is somewhat
less important; the maximum ratio of $g$ to $1/(\gamma_o - 1)$ (the
nonrelativistic value) is $1.10$.

\section{Summary}

We have written down and solved equations describing optically thin,
advection dominated flows in the Kerr metric.  Our physical
description of the flow includes a causal prescription for the
viscosity, and does not require the application of a boundary
condition at the event horizon.  We solve the energy equation assuming
that a constant fraction of the dissipated energy is advected.  We
have also included certain relativistic terms that have been neglected
in some earlier treatments; most importantly, we do not assume that
the relativistic enthalpy $\eta$ is unity.  We will present a detailed
survey of solutions for various choices of parameters in an upcoming
paper (\cite{PG}).  These solutions, when coupled with a detailed
description of the cooling processes and a photon transport scheme,
will allow us to produce model spectra for advection dominated
systems.

\acknowledgments

We are grateful to R. Narayan for helping to initiate this project and
for his support, insight and encouragement, and to J.-P. Lasota and W.
Press for helpful discussions.  This work was supported by NASA grant
NAG 5-2837 and NSF grant AST 9423209.

\appendix
\section{Shear Rate in the Local Rest Frame}

We want $\sigma_{(r)(\phi)}$, which is the shear rate
measured in the local rest frame (LRF).  Using the basis
vectors for the LRF,
\begin{equation}
\sigma_{(r)(\phi)} = {e^{\mu}}_{(r)} {e^{\nu}}_{(\phi)}
	\sigma_{\mu\nu},
\end{equation}
where $\sigma_{\mu\nu}$ is given by equation (\ref{SIGDEF}).
We must then calculate many of the pieces of $\sigma_{\mu\nu}$,
although our task is simplified a bit because
\begin{equation}
{e^{\mu}}_{(a)} u_\mu = \delta_{(a)}^{(t)},
\end{equation}
since ${e^{\mu}}_{(t)}$ is parallel to $u^\mu$, and the basis
vectors are orthonormal.  Along the way we find
\begin{eqnarray}
u_{t;t} & = & -{\beta_r\gamma_r \sqrt{\sD}\over{r^2}} \nonumber\\
u_{t;r} & = & -{d\sE\over{d r}} + u_{r;t} \nonumber \\
u_{t;\phi} & = & {a \beta_r\gamma_r \sqrt{\sD}\over{r^2}} \nonumber\\
u_{r;t} & = & (1 - a \Omega){\gamma\over{r^2}}\sqrt{{\sA\over{\sD}}}
	\nonumber\\
u_{r;r} & = & {d u_r\over{d r}} + \beta_r\gamma_r (1 - a^2/r)\\
u_{r;\phi} & = & -{\gamma\over{r^2}}\sqrt{{\sA\over{\sD}}}
	(a - a^2\Omega + r^3\Omega)\nonumber\\
u_{\phi;t} & = & u_{t;\phi}\nonumber\\
u_{\phi;r} & = & {d l\over{d r}} + u_{r;\phi}\nonumber\\
u_{\phi;\phi} & = & r\beta_r\gamma_r\sqrt{\sD} (1 - a^2/r^3).\nonumber
\end{eqnarray}
Using equation (\ref{COVVEL}), 
\begin{eqnarray}
{d\sE\over{d r}} & = & {V \gamma_r^4\over{\gamma}}\sqrt{\sD\over{\sA}}
	{d V\over{d r}} +
	\Omega {d l\over{d r}} -
	{l\omega\over{r \sA}} (3 + a^2/r^2) +
	{\gamma (1 - a^2/r)\over{r^2 \sqrt{\sA\sD}}} + \nonumber\\
	& & {\gamma\sqrt{\sD} a^2 (1 + 3/r)\over{r^3 \sA^{3/2}}} -
	{l^2 (1 - a^2/r^3)\over{r^3 \sA^2\gamma}}\sqrt{\sD\over{\sA}}
\end{eqnarray}
Assembling the results and transforming to the LRF, we find
\begin{eqnarray}
\sigma_{(r)(\phi)} & = &
	-{l V \gamma_r^4\over{2 r \gamma}}\sqrt{\sD\over{\sA}}
	{d V\over{d r}} +
	{\sE - l\Omega\over{2 r}}{d l\over{d r}} 
	 - {l^2\omega (3 + a^2/r^2)\over{2 r^2\gamma_r^2 \sA}} -
	{1\over{2}}\gamma_\phi^2 \omega (3 + a^2/r^2) \nonumber\\
	& & + {\gamma_\phi l\over{\gamma_r}}\sqrt{\sA\over{\sD}}
	\left(
		{1\over{r^3}} -
		{\sD (1 - a^2/r^3)\over{\sA^2 r^2}} -
		\omega^2 \left(2 + a^2/r^2 + a^2/r^3\right)
	\right) \\
	& & + {l^2 \omega (3 + a^2/r^2)\over{2 r^2 \sA}} -
	{l \gamma (1 - a^2/r)\over{2 r^3\sqrt{\sA\sD}}} -
	{l \gamma \sqrt{\sD} a^2 (1 + 3/r)\over{2 r^4 \sA^{3/2}}} +
	{l^3 (1 - a^2/r^3)\over{2 r^4 \sA^2\gamma}}\sqrt{\sD
		\over{\sA}}. \nonumber
\end{eqnarray}
For a relativistic thin disk, this reduces to the remarkably simple
expression
\begin{equation}\label{TDSHEAR}
\sigma_{(r)(\phi)} = {1\over{2}} r \sA \gamma_\phi^2 
	{d\Omega\over{d r}} \qquad\qquad ({\rm THIN \,\,DISK})
\end{equation}
(\cite{nt73}).  Obviously this reduces correctly to the nonrelativistic
limit.  Remarkably, this thin disk expression for $\sigma$ lies
within a fraction of a percent of the full LRF shear stress in our 
advection dominated solutions.

For cold, geodesic flow ($l = const.$, $\sE = const.$) onto a
nonrotating hole, $\sigma$ reduces to
\begin{equation}
\sigma_{(r)(\phi)} = {\gamma_\phi l \over{\gamma_r
	\sqrt{\sD}} r^2}(1 - 3/r). \qquad\qquad(a = 0, {\rm GEODESIC})
\end{equation}
This case is simple enough to be checked by hand.  It shows the
shear reversal at $r = 3$ discovered by \cite{al88} and discussed 
in detail by \cite{ap90}.

\clearpage

\begin{figure}
\plotone{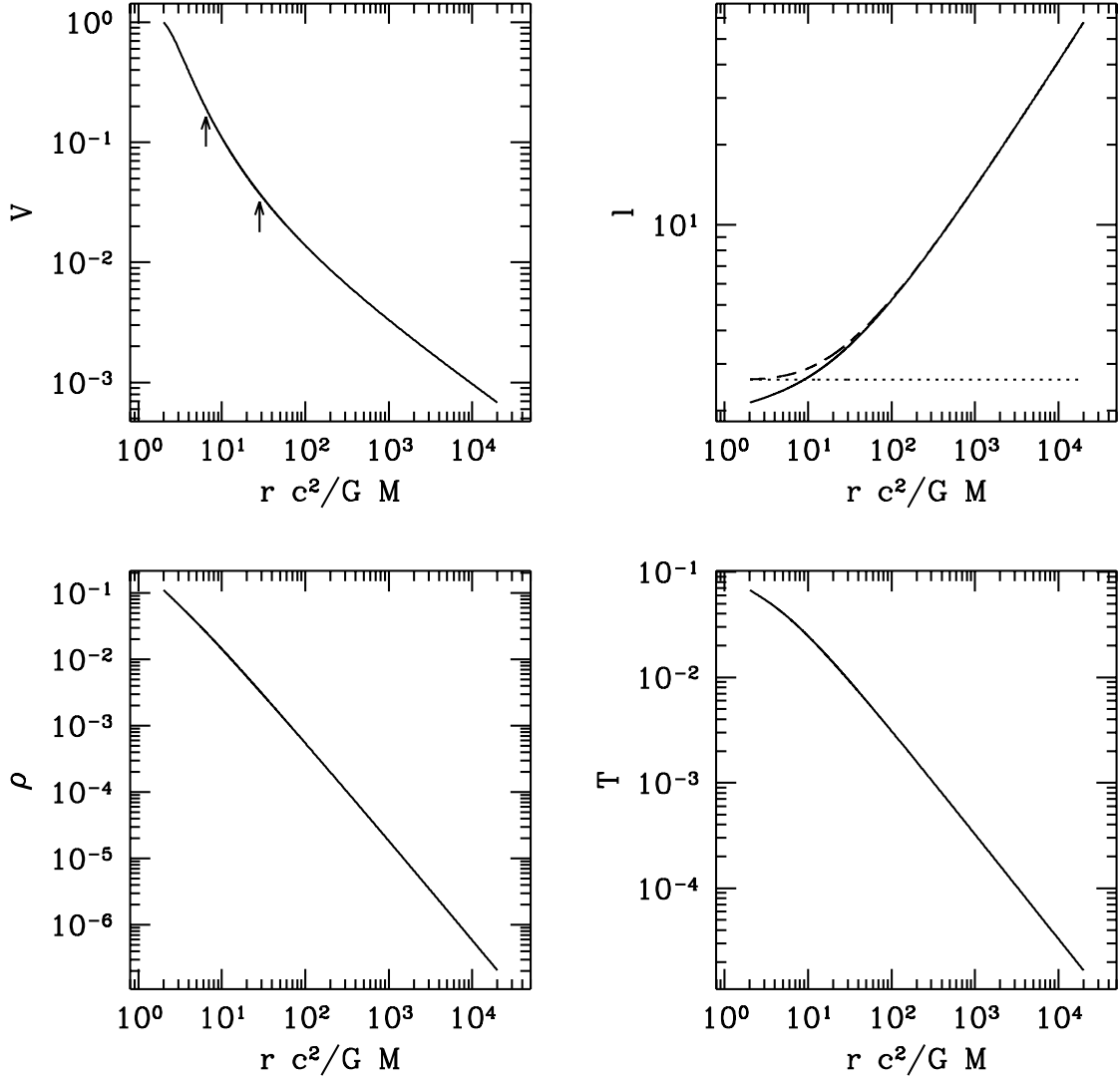}
\caption{
Solution to the full relativistic equations with $a = 0$,
$\alpha = 0.1$, $f = 1$.  The upper left panel shows the 
radial velocity $V$ measured in the CRF.  The arrows show the
location of the sonic and viscous critical points.  The
upper right panel shows the angular momentum $l$ (solid line)
as well as $l\eta$ (dashed line) and $j$ (dotted line).
The lower left panel shows density and the lower right
panel shows dimensionless temperature $T$.
}
\end{figure}

\begin{figure}
\plotone{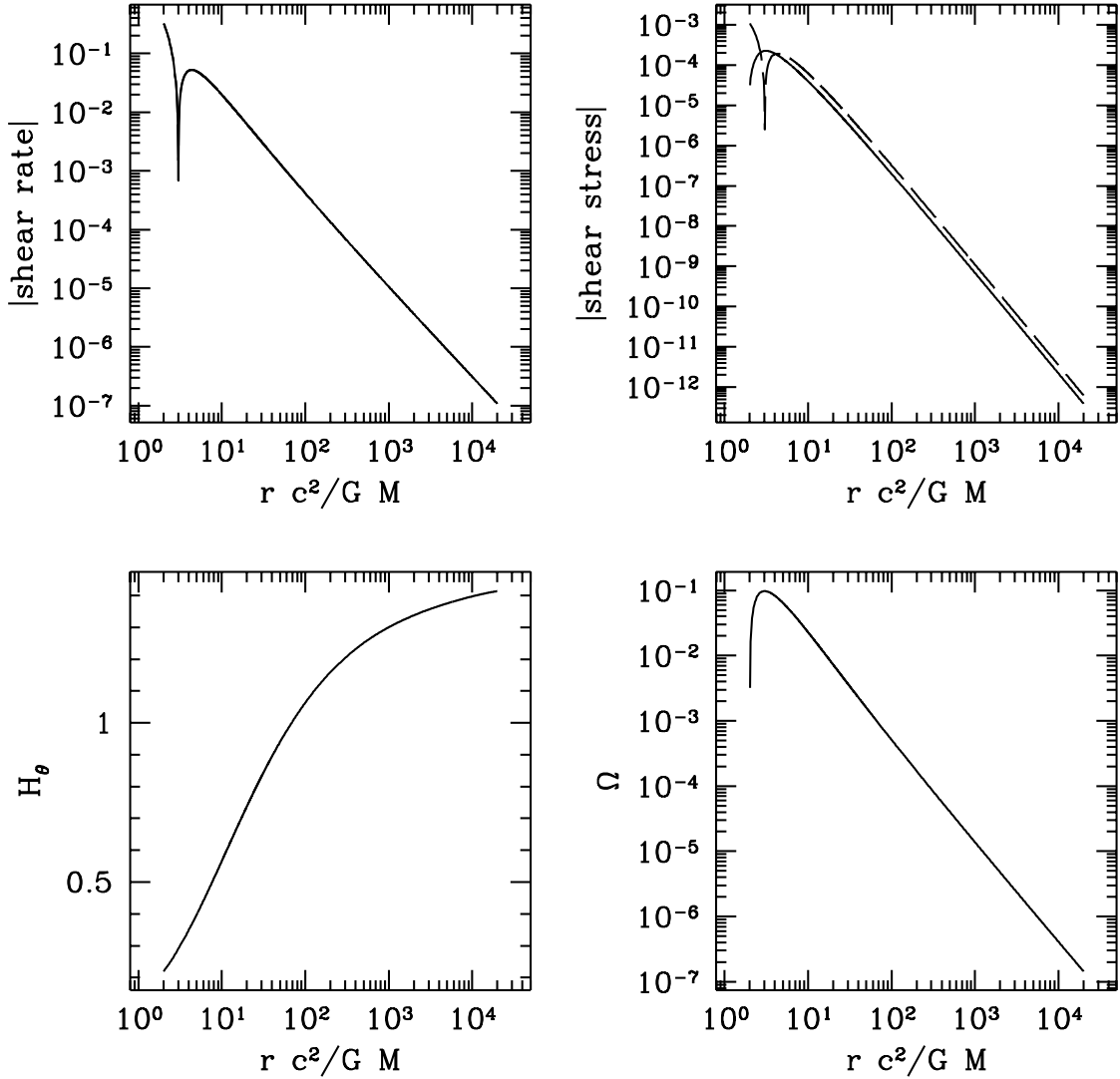}
\caption{
Dynamical quantities in the $a = 0$ solution.  The upper
left panel shows the shear rate measured in the LRF.  The
approximate formula for the shear rate, based on the thin
disk shear, is also shown but is too close to distinguish
on this plot.  The
upper right panel shows both the causal shear stress (solid
line) and the acausal (Navier-Stokes) version, which is
proportional to the shear rate (dashed line).  The lower
left panel shows the angular scale of the flow $H_\theta$, 
while the lower right shows the angular frequency $\Omega
= u^\phi/u^t$.
}
\end{figure}

\begin{figure}
\plotone{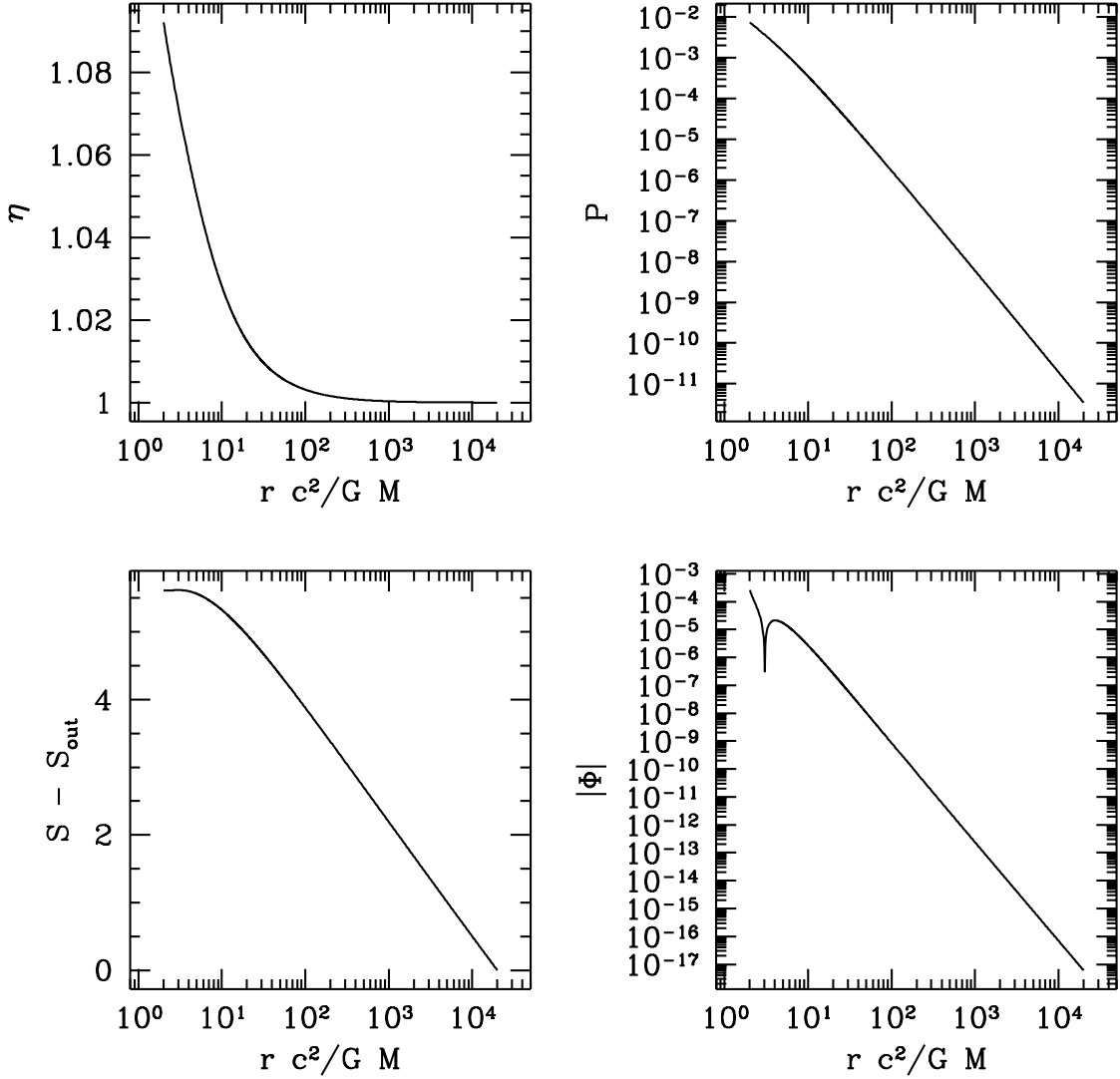}
\caption{
Thermodynamic quantities in the $a = 0$ solution.  The
upper left panel shows the relativistic enthalpy $\eta$,
the upper right panel shows pressure, the lower left
entropy, and the lower right the absolute value of the
dissipation function.  The dissipation function changes
sign at $r \simeq 3$.
}
\end{figure}

\begin{figure}
\plotone{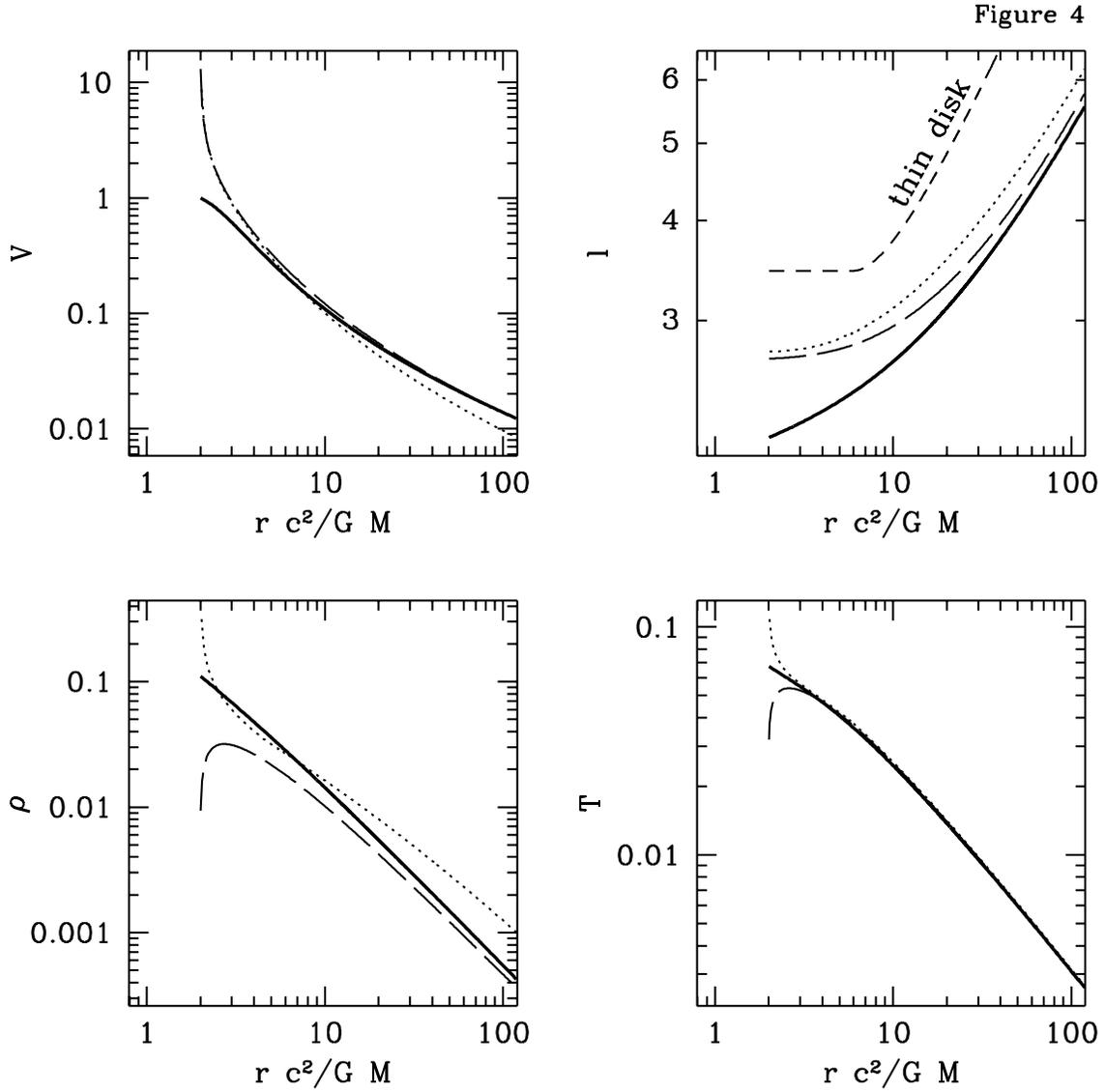}
\caption{
A comparison of various ADAF solutions: the relativistic
solution (heavy solid line); a solution of the causal
nonrelativistic limit of the relativistic equations
(long dashed line); the NKH solution (dotted line); and
the thin disk angular momentum distribution (short
dashed line).
}
\end{figure}

\begin{figure}
\plotone{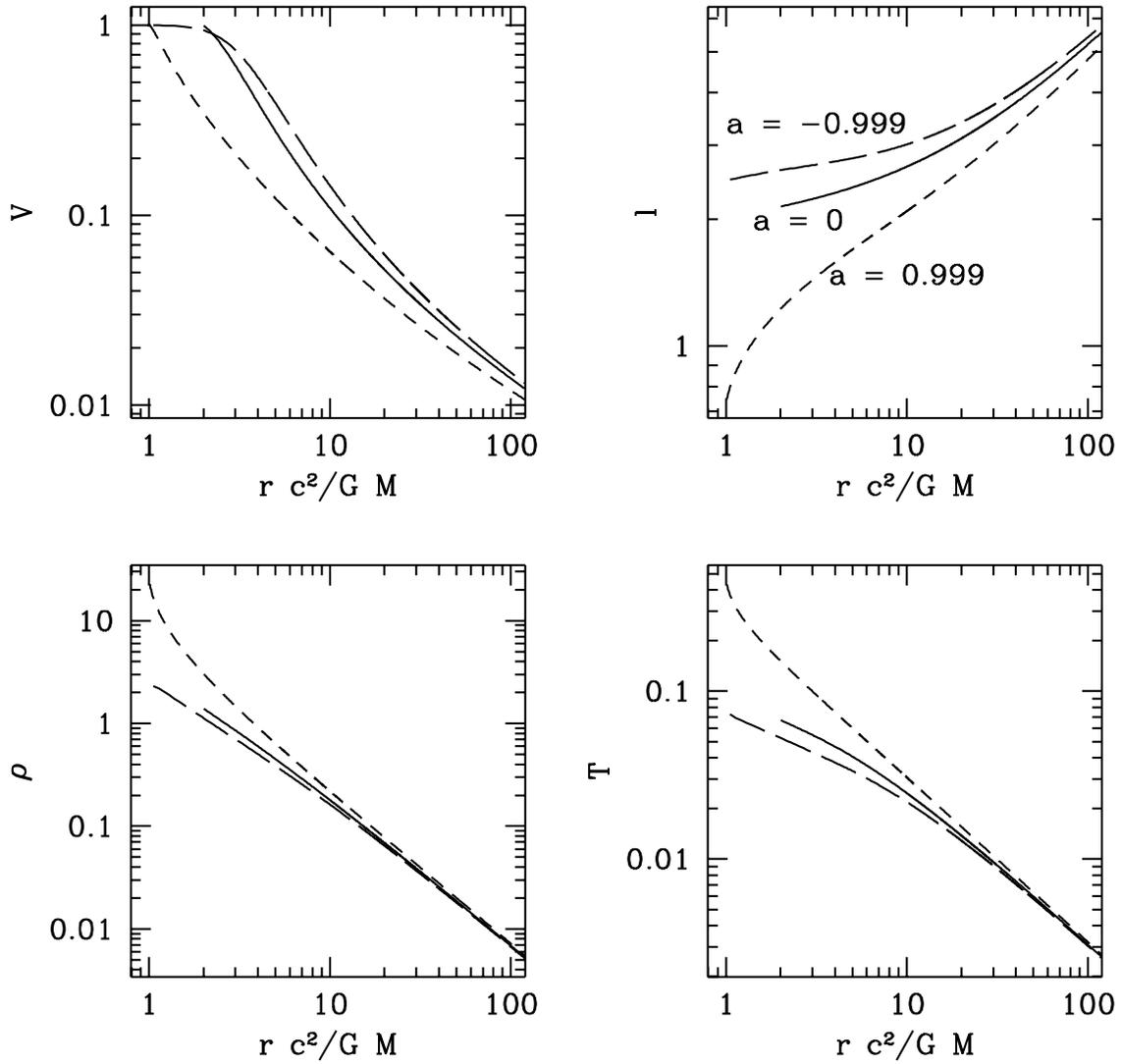}
\caption{
A comparison of the relativistic solution for different
black hole angular momenta.  Shown are solutions for
$a = 0$ (solid line), $a = -0.999$ (long dashed line)
and $a = 0.999$ (short dashed line).
}
\end{figure}

\end{document}